\documentclass[showpacs,aps,prb,twocolumn,superscriptaddress]{revtex4-1}
\usepackage[latin1]{inputenc}
\usepackage{graphicx}
\usepackage{hyperref}
\usepackage{amssymb}
\usepackage{amsmath}
\usepackage{xspace}
\usepackage{dcolumn}
\usepackage{bm}
\usepackage{color}
\usepackage{float}
\usepackage{wasysym}
\usepackage{soul}

\usepackage[extra]{tipa}

\setcitestyle{square,numbers,compress}

\newcommand{\ie}[0]{i.e.\@\xspace}

\newcommand{\ve}[1]{\boldsymbol{#1}}

\renewcommand{\tilde}[1]{\widetilde{#1}}
\newcommand{\tr}[0]{\text{tr}}

\begin{document}

%%%%%%%%%%%%%%%%%%%%%%%%%%%%%%%%%%%%%%%%%%%%%%%%%%%%%%%%%%%%%%%%%%%%%%%%%%%%%
%%%%%%%%%%%%%%%%%%%%% TITLE & ABSTRACT %%%%%%%%%%%%%%%%%%%%%%%
%%%%%%%%%%%%%%%%%%%%%%%%%%%%%%%%%%%%%%%%%%%%%%%%%%%%%%%%%%%%%%%%%%%%%%%%%%%%%

\title{Thermodynamic and dynamical signatures\\ of a quantum spin-Hall insulator to superconductor transition}

\author{Martin Hohenadler}\altaffiliation{These authors contributed equally
  to this work and are listed alphabetically.}
\affiliation{\mbox{Institut f\"ur Theoretische Physik und Astrophysik,
    Universit\"at W\"urzburg, 97074 W\"urzburg, Germany}}
\affiliation{Independent Researcher, Josef-Retzer-Str.~7, 81241 Munich, Germany}
\author{Yuhai Liu}\altaffiliation{These authors contributed equally
  to this work and are listed alphabetically.}
\affiliation{\mbox{Beijing Computational Science Research Center, 10 East Xibeiwang Road, Beijing 100193, China}}
\author{Toshihiro Sato}\altaffiliation{These authors contributed equally
  to this work and are listed alphabetically.}
\affiliation{\mbox{Institut f\"ur Theoretische Physik und Astrophysik, Universit\"at W\"urzburg, 97074 W\"urzburg, Germany}}
\author{Zhenjiu Wang}
\affiliation{\mbox{Institut f\"ur Theoretische Physik und Astrophysik, Universit\"at W\"urzburg, 97074 W\"urzburg, Germany}}
\affiliation{\mbox{Max Planck Institute for the Physics of Complex Systems, N\"othnitzerstr. 38, 01187 Dresden, Germany}}
\author{Wenan Guo}
\affiliation{\mbox{Department of Physics, Beijing Normal University, Beijing 100875, China }}
\affiliation{\mbox{Beijing Computational Science Research Center, 10 East Xibeiwang Road, Beijing 100193, China}}
\author{Fakher F. Assaad} \email{fakher.assaad@physik.uni-wuerzburg.de}
\affiliation{\mbox{Institut f\"ur Theoretische Physik und Astrophysik, Universit\"at W\"urzburg, 97074 W\"urzburg, Germany}}
\affiliation{\mbox{W\"urzburg-Dresden Cluster of Excellence ct.qmat, Am Hubland, 97074 W\"urzburg, Germany}}

\begin{abstract}
  Thermodynamic and dynamical properties of a model of Dirac fermions with a
  deconfined quantum critical point (DQCP) separating an interaction-generated
  quantum spin-Hall insulator from an s-wave superconductor [Nature Comm.~{\bf
    10}, 2658 (2019)] are studied by quantum Monte Carlo simulations.  Inside
  the deconfined quantum critical region bound by the single-particle gap,
  spinons and spinless charge-2e skyrmions emerge. Since the model
  conserves total spin and charge, and has a single length scale, these
  excitations lead to a characteristic linear temperature dependence of the
  uniform spin and charge susceptibilities. At the DQCP, the order parameter
  dynamic structure factors show remarkable similarities that support emergent Lorentz 
  symmetry. Above a critical temperature, superconductivity is destroyed by the
  proliferation of spin-1/2 vortices.
 \end{abstract}

\maketitle

\section{Introduction}\label{sec:introduction}

Phase transitions involving spontaneous symmetry breaking
are a central topic of condensed matter physics. Zero-temperature ($T=0$) {\it
  quantum phase transitions} \cite{Sachdev2011}, driven purely by quantum
fluctuations, are particularly interesting. The tuning of external parameters
such as pressure in their experimental realization can be thought of as changing
the relative strength of different terms in the relevant microscopic Hamiltonian
or field theory, thereby favoring distinct ground states above and below a
critical value $\lambda_c$. While theoretical approaches often focus on $T=0$,
experiments rely on the finite-temperature signatures of such quantum critical
points (QCPs) \cite{Gegenwart08}. In the quantum critical region $k_\text{B} T > \Delta$, where
$\Delta \sim \xi^{-z} \sim |\lambda-\lambda_c|^{z\nu}$ is a $T=0$ energy scale
and $z$, $\nu$ are critical exponents, quantum {\it and} thermal fluctuations
play a crucial role \cite{Sachdev2011}.

Deconfined QCPs (DQCPs) \cite{Senthil04_1,Senthil04_2}
separate two phases with different order parameters. The paradigmatic example is the transition from an
antiferromagnet (AFM) to a valence bond solid (VBS). Within a Ginzburg-Landau description, a
continuous transition between two different orders requires fine
tuning. Otherwise, either a coexistence region or a first-order transition is
expected. The CP$^1$ theory of DQCPs \cite{Senthil04_1,Senthil04_2},
provides a generic mechanism for a continuous transition by means of
topological defects of the order parameters. Vortex defects
of the VBS order parameter carry spin-$1/2$.  These  spinons are  the 
fundamental  emergent  quantity of the  theory.  At the  critical 
point, they are  deconfined  and  form  a  U(1)  spin liquid  \cite{Xu18}. The AFM phase 
arises  from spinon  condensation. In contrast, in the VBS phase, the spinons are
gapped and VBS order is associated with the condensation of monopoles of the
gauge field \cite{Polyakov_book}. The latter are equivalent to skyrmions of the
AFM order parameter and carry C$_4$ charge.

\begin{figure}[t]
  \centering
  \includegraphics[width=0.48\textwidth]{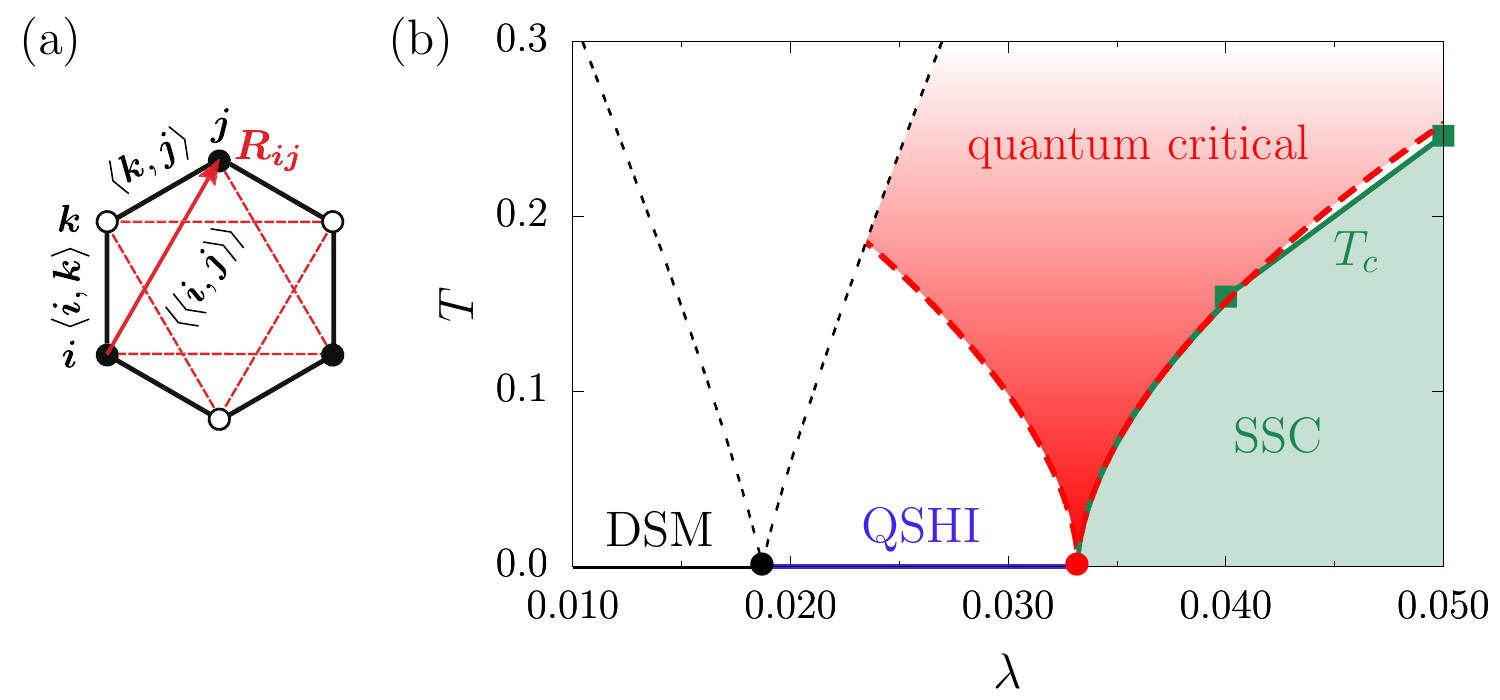}
  \caption{\label{fig:phasediagram}
    (a) Notation in Eq.~(\ref{eq:hamiltonian1}).
    (b) Phase diagram of
    Hamiltonian~(\ref{eq:hamiltonian1}). Two quantum critical points at
    $\lambda_{c1}=0.0187(2)$ [Dirac semimetal (DSM) to quantum spin Hall
    insulator (QSHI)] and $\lambda_{c2}=0.0332(2)$ [QSHI to s-wave
    superconductor (SSC)] were established in
    Refs.~\cite{liu2019superconductivity,PhysRevB.104.035107}. The DQCP quantum
    critical region is shown above $T=5.35|\lambda-\lambda_{c2}|^{z\nu}$ (red
    dashed lines, $z=1$, $\nu=0.58$, prefactor obtained using
    $T_c(\lambda=0.04)\approx1/6.5$) and to the right of the crossover
    temperature scale $30.7|\lambda-\lambda_{c1}|^{z\nu}$ associated with the
    DSM-QSHI QCP (black dashed lines, $z=1$, $\nu=1/1.14$
    \cite{liu2019superconductivity}, prefactor from
    $\Delta_\text{sp}(\lambda=0.026)\approx 0.41$). The curve for $T_c(\lambda)$
    corresponds to $5.35|\lambda-\lambda_{c2}|^{z\nu}$ for $\lambda\leq0.04$ and
    a straight line for $\lambda>0.04$.  }
\end{figure}

Here, we investigate finite-temperature as well as spectral signatures
of a DQCP within a recently introduced fermionic model
\cite{liu2019superconductivity} whose phase diagram is shown in
Fig.~\ref{fig:phasediagram}(b).  It describes interacting Dirac fermions in 2+1
dimensions and can be studied by quantum Monte Carlo (QMC) simulations without a
sign problem. The DQCP separates an interaction-generated quantum spin-Hall
insulator (QSHI) from an s-wave superconductor (SSC). Whereas the field theory
description of the DQCP is based on an SO(3)$\times$U(1) symmetry, the U(1)
component is merely emergent in AFM-VBS lattice realizations. In contrast, the
current QSH-SSC case has an exact U(1) symmetry at the
Hamiltonian level corresponding to global charge conservation.
This has important consequences: skyrmions of the SO(3) QSH order
parameter carry charge 2e \cite{Grover08} and are therefore conserved.  On the
other hand, spinons corresponding to vortices of the SSC order parameter have
spin-1/2 \cite{Grover08}.  The thermodynamic and dynamical response of the
system inside the quantum critical fan will be dominated by
topology since charge and spin responses capture skyrmion and spinon degrees of
freedom, respectively. In contrast to previous work on finite-$T$ quantum
criticality of AFM-VBS DQCPs
\cite{PhysRevLett.100.017203,Sandvik11,PhysRevB.87.180404}, our 
fermionic approach allows us to characterize the topological excitations of the
state inside the critical fan with susceptibilities of conserved
quantities. Rather uniquely, it has gapless spin and  charge excitations but a
nonzero single-particle gap.

The remainder of the paper is organized as follows. The model is introduced in
Sec.~\ref{sec:model}, followed by a discussion of the methods used in
Sec.~\ref{sec:method}. Results are presented in Sec.~\ref{sec:results} and
discussed in Sec.~\ref{sec:discussion}. We provide an appendix on the scaling of
conserved quantities.

\section{Model}\label{sec:model}

We consider the Hamiltonian \cite{liu2019superconductivity}
\begin{eqnarray}
\label{eq:hamiltonian1}
\hat{H} =
  -t\sum_{\langle \ve{i}, \ve{j} \rangle}\left(\hat{\boldsymbol{c}}^{\dag}_{\ve{i}} \hat{\boldsymbol{c}}^{}_{\ve{j}}+\text{H.c.}\right)
  -\lambda\sum_{\hexagon} \Bigg( \sum_{ \langle \langle \ve{i},\ve{j} \rangle \rangle  \in \hexagon }\hat{J}_{\bm{i},\bm{j}} \Bigg)^2,\nonumber \\
\end{eqnarray}
where
$ \hat{J}_{\bm{i},\bm{j}} = i \nu_{ \bm{i} \bm{j} }
\hat{\ve{c}}^{\dagger}_{\bm{i}} \bm{\sigma}
\hat{\ve{c}}^{\phantom\dagger}_{\bm{j}} + \text{H.c.}$
We use the spinor notation $\hat{\boldsymbol{c}}^{\dag}_{\ve{i}} =
\big(\hat{c}^{\dag}_{\ve{i},\uparrow},\hat{c}^{\dag}_{\ve{i},\downarrow}
\big)$; $\hat{c}^{\dag}_{\ve{i},\sigma} $ creates an electron at lattice
site $\ve{i}$ on a honeycomb lattice (see also Fig.~\ref{fig:phasediagram}(a))
with physical spin $\sigma$. The first term in Eq.~(\ref{eq:hamiltonian1})
accounts for nearest-neighbor hopping, the second term is a plaquette interaction
involving next-nearest-neighbor pairs of sites and phase factors
$\nu_{\boldsymbol{ij}}=\pm1$ identical to those of the Kane-Mele model \cite{KaneMele05b}.
The components of $\boldsymbol{\sigma}=(\sigma^x,\sigma^y,\sigma^z)$ are the Pauli spin-1/2 matrices.
We consider half-filling and work in units where $\hbar=k_\text{B}=t=1$.

Figure~\ref{fig:phasediagram}(b) shows the phase diagram. The $T=0$ line was
previously established in
Refs.~\cite{liu2019superconductivity,PhysRevB.104.035107}.  The interaction is
irrelevant for $\lambda<\lambda_{c1}=0.0187(2)$, where the ground state is a
Dirac semimetal (DSM). For $\lambda_{c1}<\lambda<\lambda_{c2}=0.0332(2)$, the
SU(2) spin symmetry of Eq.~(\ref{eq:hamiltonian1}) is spontaneously broken by
long-range QSH order. For $\lambda>\lambda_{c2}$, the spin symmetry is restored
and the QSH order gives way to the SSC with a spontaneously broken U(1)
symmetry. The DSM-QSHI transition exhibits fermionic Gross-Neveu-Heisenberg
criticality \cite{liu2019superconductivity,PhysRevB.104.035107}, whereas the
QSHI-SSC transition appears to be described by the theory of DQCPs.  Similar to
AFM-VBS transitions in other fermionic models
\cite{SatoT17,Li17,PhysRevB.98.121406}, the QSHI-SSC transition can be
understood within a bosonic (single-fermion excitations being gapped) DQCP
picture: the QSHI is destroyed by the proliferation of charge-2e skyrmion
defects in its order parameter whose condensation produces the SSC
\cite{Grover08}. At $T>0$, long-range QSH order is prohibited by the
Mermin-Wagner theorem, leading to a finite-temperature crossover (dashed line
emerging from the DQCP in Fig.~\ref{fig:phasediagram}(b)).  The SSC will be shown to undergo a
Berezinskii-Kosterlitz-Thouless phase transition at $T_c(\lambda)>0$ (solid
line). The dashed lines in Fig.~\ref{fig:phasediagram}(b) provide an estimate
of the range of the quantum critical region. The latter is not defined by a
sharp transition line but by a crossover \cite{Sachdev2011}. However, a temperature scale can be
derived from the known scaling functions as well as the single-particle gap (for
$\lambda<\lambda_{c2}$) and $T_c$ (for $\lambda>\lambda_{c2}$), respectively
(see Fig.~\ref{fig:phasediagram} for details).

In the following, we focus on the couplings $\lambda=0.026$ (QSHI), $\lambda=0.0332$ (DQCP), and
$\lambda=0.050$ (SSC). As shown in Fig.~\ref{fig:sp-green} and discussed below,
over this range, Hamiltonian~(\ref{eq:hamiltonian1}) has a nonzero single-particle gap
$\Delta_\text{sp}$, which has important consequences for its symmetries.
The generator of the model's SU(2) spin symmetry is the total spin, $\hat{\ve{S}}_{\text{tot}} = \frac{1}{2} \sum_{\ve{i}, \sigma, \sigma'}
\hat{c}^{\dagger}_{\ve{i},\sigma} \ve{\sigma}_{\sigma \sigma'}
\hat{c}^{}_{\ve{i},\sigma'} $.  Under a $2\pi$ rotation, $e^{2 \pi i \ve{e}\cdot \hat{\ve{S}}_{\text{tot}} } = (-1)^{\hat{N}_{tot}}$,
where $\hat{N}_{\text{tot}} = \sum_{\ve{i}, \sigma}
\hat{c}^{\dagger}_{\ve{i},\sigma} \hat{c}^{}_{\ve{i},\sigma} $ corresponds to
the total charge and $\ve{e}$ is a unit vector in $\mathbb{R}^3$.  The nonzero
gap pins the parity of the low-energy sector in which
$2\pi$ rotations leave all quantities invariant.  In other words, our model
possesses a low-energy SO(3)$\times$U(1) symmetry where U(1) corresponds to
charge conservation  \cite{Thorngren20}.

\section{Methods}\label{sec:method}

The symmetry of Hamiltonian~(\ref{eq:hamiltonian1}) under time reversal ensures
the absence of a sign problem in our QMC simulations \cite{Wu04,liu2019superconductivity}.

To study finite-temperature properties, we used the ALF (Algorithms
for Lattice Fermions) implementation \cite{ALF_v1} of the finite-temperature, auxiliary-field QMC
method \cite{Blankenbecler81,Assaad08_rev}.  We simulated lattices with
$L \times L$ unit cells (each containing two orbitals) using a Trotter
discretization $\Delta_\tau=0.2$ and periodic boundary conditions.

Results for dynamical quantities  were obtained at $T=0$ with the projective QMC
algorithm from the
ALF library \cite{alfcollaboration2021alf}. This canonical algorithm filters
out the ground state  $|\psi_0\rangle $ from a trial wave function $|\psi_T
\rangle$, required to be non-orthogonal to $|\psi_0\rangle $,
\begin{equation}
 \frac{  \langle \psi_0 |  \hat{O} | \psi_0  \rangle   }{\langle \psi_0 |   \psi_0  \rangle } = \lim_{\Theta \rightarrow \infty}
  \frac{ \langle \psi_T |  e^{- \Theta \hat{H}}  \hat{O}  e^{- \Theta \hat{H}}  | \psi_T \rangle }{
   \langle \psi_T |  e^{ - 2\Theta \hat{H} } | \psi_T \rangle   }.
\end{equation}
We made the same choice for the Slater determinant for $|\psi_0\rangle $ as in
Ref.~\cite{wang2020dopinginduced}, breaking lattice and point group
symmetries but preserving time-reversal symmetry. Since $\lambda > 0$, we can decouple
the  interaction in Eq.~(1) with a real-valued Hubbard-Stratonovich
transformation. This makes both the imaginary-time propagation and the trial
wave function invariant under time reversal. Consequently, the eigenvalues of
the fermion matrix come in complex conjugate pairs and no sign problem arises.

We found a projection length $\Theta =L $ sufficient to converge to
the finite-size ground state for the system sizes considered and used a symmetric
Trotter decomposition to ensure Hermiticity of the imaginary time propagator;
the Trotter time step was $\Delta_{\tau}=0.2$.

At $T=0$, there is no distinction between particle, particle-hole and
particle-particle channels. Accordingly, for an operator $\hat{O}_{\ve{q}}$, the
dynamical structure factor in the Lehmann representation takes the form
\begin{equation}\label{Eq:spectral}
  C^O ( \boldsymbol{q}, \omega)  \equiv    \pi   \sum_{ n}  | \langle  \Psi_n |  \hat{O}_{\boldsymbol{q}}  | \Psi_0 \rangle |^2
  \delta( E_n - E_0 - \omega )
\end{equation}
with $ \hat{H} |\Psi_n \rangle = E_n |\Psi_n \rangle $.  The QMC simulations
yield correlators in imaginary time,
\begin{equation}\label{eq:dynamic-correlator}
  O(\ve{q},\tau)  =   \langle  \Psi_0  | \hat{O}_{\ve{q}}^{\dagger}(\tau)  \hat{O}_{\ve{q}} | \Psi_0 \rangle,
\end{equation}
which are related to the real-time observables by
\begin{equation}
  O(\boldsymbol{q}, \tau)
  = \frac{1}{\pi}
  \int  \, {\rm d} \omega \,  e^{- \tau \omega} O(\boldsymbol{q},\omega).
\end{equation}
The analytical continuation was done with the ALF implementation
\cite{alfcollaboration2021alf} of the stochastic maximum entropy method
\cite{Sandvik98,Beach04a}.

\section{Results}\label{sec:results}

\subsection{Finite-temperature properties}

To investigate temperature effects, we computed the trace  $\chi^{\alpha}(\boldsymbol{q})=\tr
\chi^{\alpha}_{\ve{\delta},\ve{\delta}'} (\boldsymbol{q})$ of the susceptibilities
\begin{eqnarray}
\label{eq:chi}
\chi^{\alpha}_{\ve{\delta},\ve{\delta}'} (\boldsymbol{q})
=\frac{1}{L^2} \sum_{\boldsymbol{r},\boldsymbol{r'}}\int_{0}^{\beta} \text{d} \tau
 e^{\mathrm{i}\boldsymbol{q}\cdot(\boldsymbol{r}-\boldsymbol{r}')}
\langle
  \hat{\boldsymbol{O}}^\alpha_{\boldsymbol{r},\boldsymbol{\delta}}(\tau)
  \hat{\boldsymbol{O}}^\alpha_{\boldsymbol{r'},\boldsymbol{\delta}'}(0)
\rangle \nonumber \\
\end{eqnarray}
for spin current operators
\begin{equation}
\hat{\boldsymbol{O}}^{\text{QSH}}_{\boldsymbol{r},\boldsymbol{\delta}} =
\mathrm{i}\hat{\boldsymbol{c}}^{\dagger}_{\boldsymbol{r}} \boldsymbol{\sigma}
\hat{\boldsymbol{c}}^{}_{\boldsymbol{r}+ \boldsymbol{\delta}} +
\text{H.c.}
\end{equation}
and s-wave paring operators
\begin{equation}
\hat{O}^\text{SSC}_{\ve{r},\ve{\tilde{\delta}}} =\frac{1}{2}\left(
\hat{c}^{\dagger}_{\ve{r} +\ve{\tilde{\delta}},\uparrow}
\hat{c}^{\dagger}_{\ve{r} +\ve{\tilde{\delta}},\downarrow} + \text{H.c.}\right)\,.
\end{equation}
Here, $\beta=1/T$, $\ve{r}$ corresponds to a hexagonal unit cell,
$\ve{r} + \ve{\delta}$ runs over all next-nearest neighbors, and
$\ve{r} +\ve{\tilde{\delta}}$ over the two orbitals of unit cell $\ve{r}$.

\begin{figure}[t]
  \centering
  \includegraphics[width=0.48\textwidth]{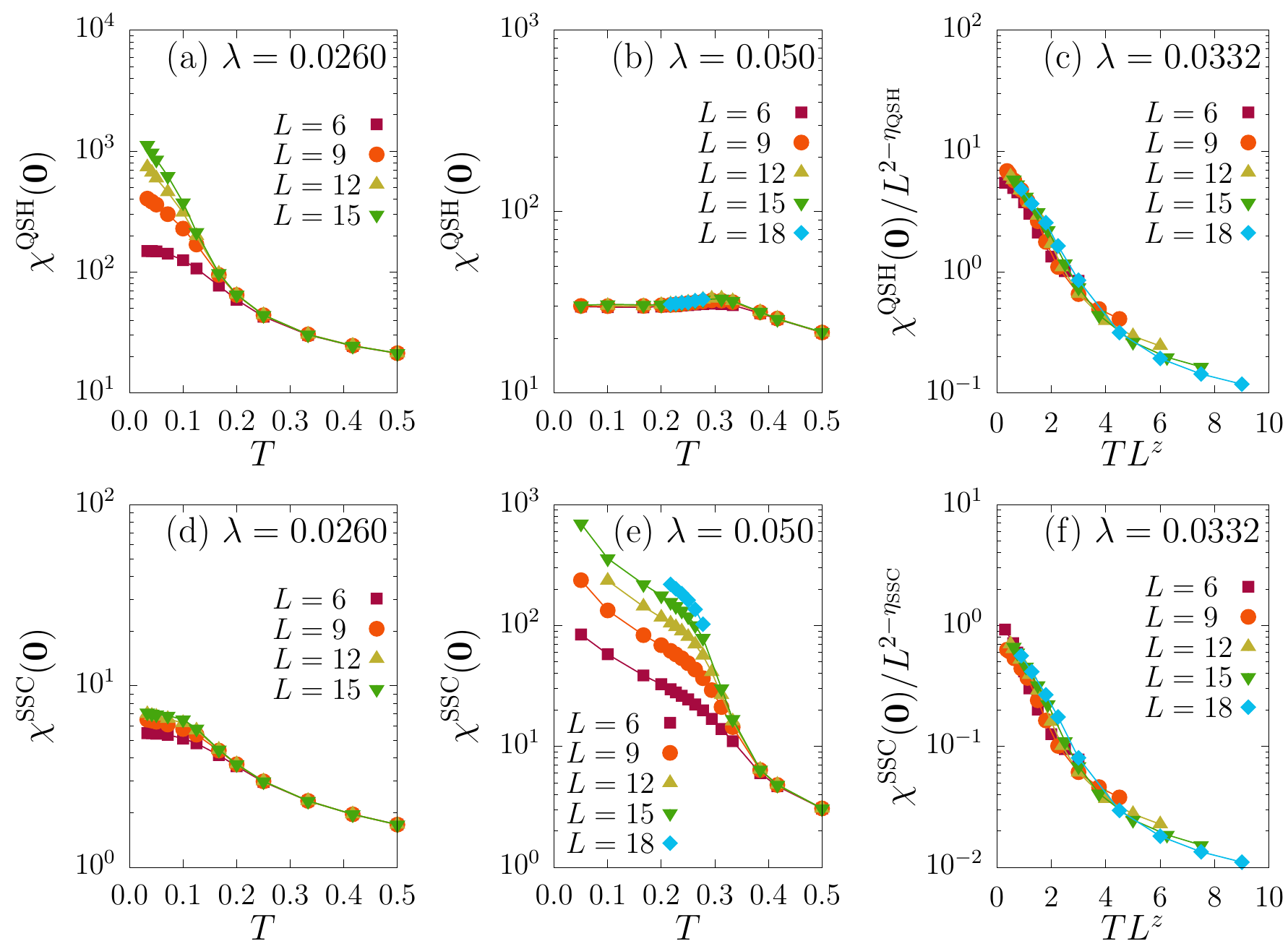}
  \caption{\label{fig:QSHSSCsus}
    (a)--(c) QSH and (d)--(f) SSC susceptibilities,
    revealing long-range QSH order at $T=0$ for $\lambda=0.026$
    [(a),(d)] and s-wave superconductivity at $T<T_c$ for $\lambda=0.050$ [(b),(e)].
    Data collapse at the DQCP using  $\lambda=\lambda_{c2}=0.0332$, $\eta_{\text{QSH}}=0.21$,
    $\eta_{\text{SSC}}=0.22$, and $z=1$ [(c),(f)].
  }
\end{figure}

\begin{figure}[t]
  \centering
 \includegraphics[width=0.35\textwidth]{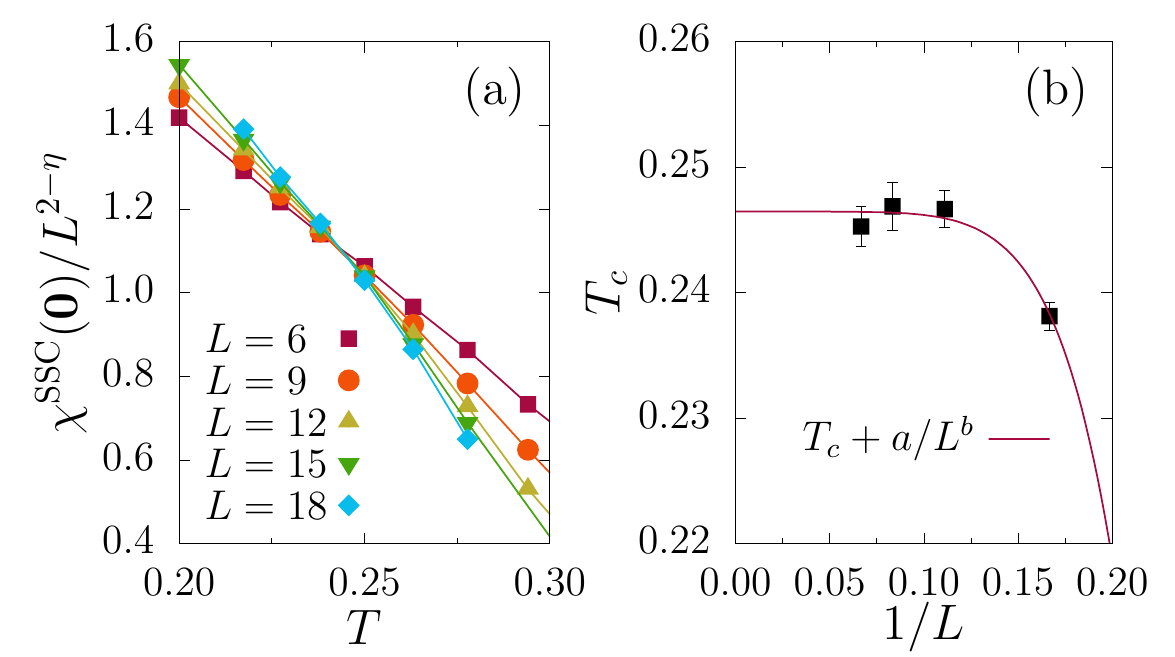}
  \caption{\label{fig:TcSSC}
    (a) Scaling analysis of the SSC susceptibility using $\eta=0.25$.
    (b) Extraction of $T_c$ from finite-size extrapolation of
     the crossing points. Here, $\lambda=0.050$ and $T_c=0.0246(1)$.}
\end{figure}

The QSHI and SSC are characterized by a divergence of the corresponding susceptibility
$\chi^{\alpha}(\boldsymbol{q}=\boldsymbol{0})$.  At $\lambda=0.026$, $\chi^\text{QSH}(\boldsymbol{0})$
increases strongly with increasing $L$ at low temperatures
(Fig.~\ref{fig:QSHSSCsus}(a)), whereas $\chi^\text{SSC}(\boldsymbol{0})$
saturates (Fig.~\ref{fig:QSHSSCsus}(d)). Conversely, at $\lambda=0.050$,
$\chi^\text{QSH}(\boldsymbol{0})$ in Fig.~\ref{fig:QSHSSCsus}(b) is essentially
independent of $L$ whereas $\chi^\text{SSC}(\boldsymbol{0})$ in
Fig.~\ref{fig:QSHSSCsus}(e) grows with increasing $L$ at low $T$.
A finite-size scaling of $\chi^\text{SSC}(\boldsymbol{0})$ followed by an
extrapolation of crossing points (Fig.~\ref{fig:TcSSC}) yields
$T_c=0.0246(1)$ for the SSC at $\lambda=0.05$, the value shown
in Fig.~\ref{fig:phasediagram}(b).

At criticality, and given Lorentz invariance \cite{Senthil04_2,PhysRevLett.100.017203}, we expect
$\chi^{\alpha}(\boldsymbol{0}) = L^{(2-\eta^{\alpha})}
f(\beta/L)$ \cite{Vicari14}, consistent with the results for $\lambda=\lambda_{c2}=0.0332$
in Figs.~\ref{fig:QSHSSCsus}(c) and (f). Using the exponents $\eta^{\text{QSH}}=0.21(5)$ and
$\eta^{\text{SSC}}=0.22(6)$ from Ref.~\cite{liu2019superconductivity},
the data collapse has no free parameters.

Hamiltonian~(\ref{eq:hamiltonian1}) has global SU(2) spin and U(1) charge
symmetry. For conserved quantities, the uniform
susceptibilities correspond to the fluctuations of the symmetry generators
(here, $\ve{\hat{G}}^{c} =  \hat{N}_{\text{tot}}$,  $\ve{\hat{G}}^{s} =
\hat{\ve{S}}_{\text{tot}}$),
\begin{equation}
\chi^{\alpha}= \frac{\beta }{L^d}  \left(   \left< \ve{\hat{G}}^{\alpha} \cdot \ve{\hat{G}}^{\alpha} \right>  -
\left< \ve{\hat{G}}^{\alpha} \right> \cdot \left< \ve{\hat{G}}^{\alpha} \right>   \right).
\end{equation}

\begin{figure}[t]
  \centering
  \includegraphics[width=0.48\textwidth]{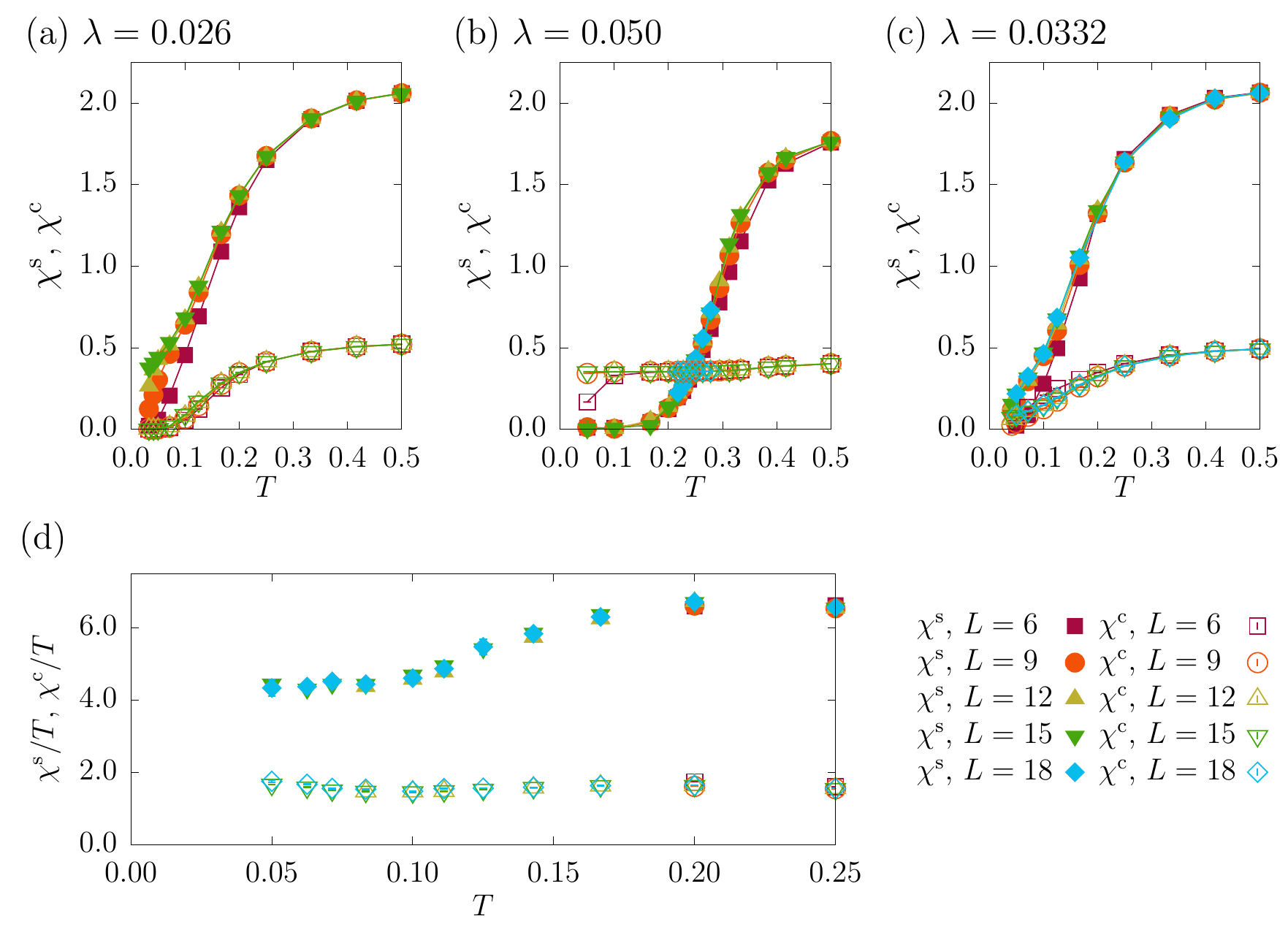}
  \caption{\label{fig:spinchargesus}
   (a)--(c) Temperature dependence of the uniform spin and charge
   susceptibilities. (d) Rescaled susceptibilities for selected parameters (see text), revealing linear behavior at
   low temperatures. The key applies to all panels.}
\end{figure}

In the QSHI, Fig.~\ref{fig:spinchargesus}(a), $\chi^{\text{s}}$ scales to a
finite value related to spin-wave excitations of the order parameter, whereas
$\chi^{\text{c}}$  vanishes exponentially for $T\to 0$, reflecting insulating behavior.
In the SSC, Fig.~\ref{fig:spinchargesus}(b), spins are entangled in pairs,
leading to an exponential decay of $\chi^{\text{s}}$. At the same time, 
because a superconductor has no gap for adding pairs, the corresponding charge
susceptibility remains finite. At the critical point, Fig.~\ref{fig:spinchargesus}(c),
given a single length scale, susceptibilities of conserved quantities follow a
hyper-scaling law. As shown in the Appendix, for $\beta = L^z$, $\chi \propto \beta^{{1 -{d}/{z}}}$.
Here, we have $d=2$ and expect $z=1$, which implies a linear temperature dependence of
$\chi^{\text{s}}$ and $\chi^{\text{c}}$. Retaining only size-converged data points---and keeping in
 mind potential, small deviations of $\lambda=0.0332$ from the
exact critical value $\lambda_{c2}$---the rescaled susceptibilities $\chi^{\text{s}}/T$ and
$\chi^{\text{c}}/T$ in Fig.~\ref{fig:spinchargesus}(d) approach a constant at low
temperatures to acceptable accuracy, consistent with linear susceptibilities.

\subsection{Dynamical properties}

\subsubsection{Single-particle excitations}

\begin{figure}[b]
  \centering \includegraphics[width=0.4\textwidth]{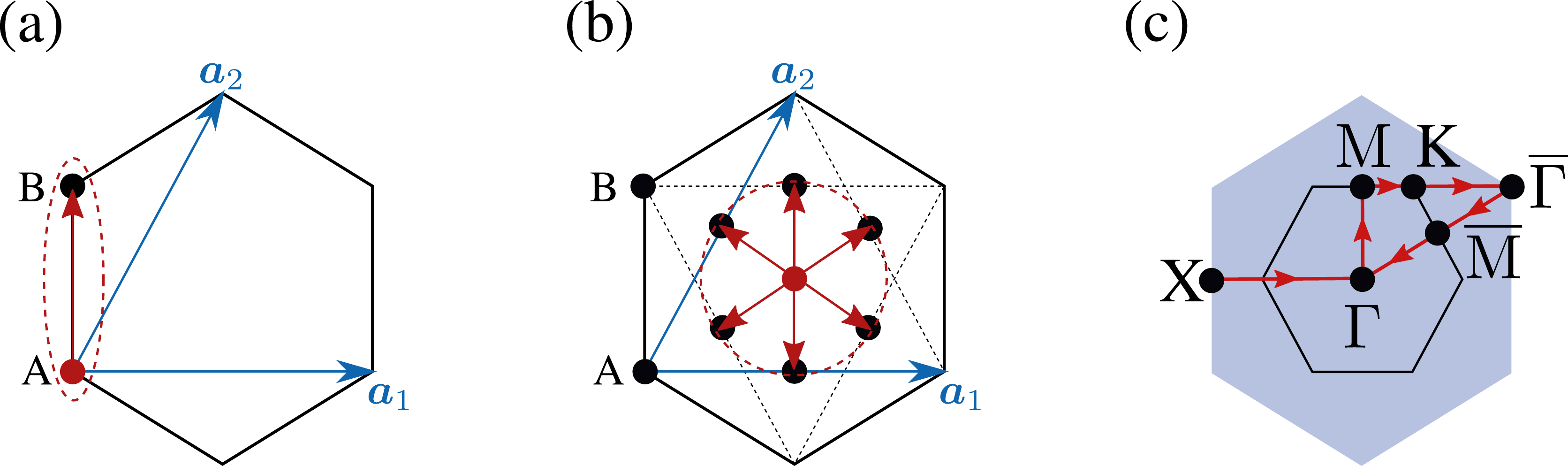}
  \caption{\label{fig:extended-zone} Notation used in the context of the
    extended zone scheme and illustration for (a) local operators and
    (b) nonlocal QSH operators. (c) Path in the Brillouin zone.}
\end{figure}

Dynamical properties where calculated in an extended zone
scheme in which $\hat{O}_{\ve{q}} $ for Eq.~(\ref{eq:dynamic-correlator}) is given by
\begin{eqnarray}
  \label{cq.eq}
  \hat{c}_{\ve{q},\sigma}  = \frac{1}{\sqrt{V}} \sum_{\ve{r}}
  e^{i  \ve{q} \cdot \ve{r}} \left(  \hat{a}_{\ve{r},\sigma} + \hat{b}_{\ve{r},\sigma}e^{i\ve{q} \cdot \ve{R}}   \right)
\end{eqnarray}
with $\ve{R}=2( \boldsymbol{a}_2-\boldsymbol{a}_1/2)/3$. Here, $\ve{r}$ runs
over unit cells and $ \hat{a}_{\ve{r},\sigma} $, $ \hat{b}_{\ve{r},\sigma} $ are
fermion annihilation operators of Wannier states centered around the A and B
sites of the unit cell, as illustrated in Fig.~\ref{fig:extended-zone}(a).

\begin{figure}[t]
  \centering
  \includegraphics[width=0.48\textwidth]{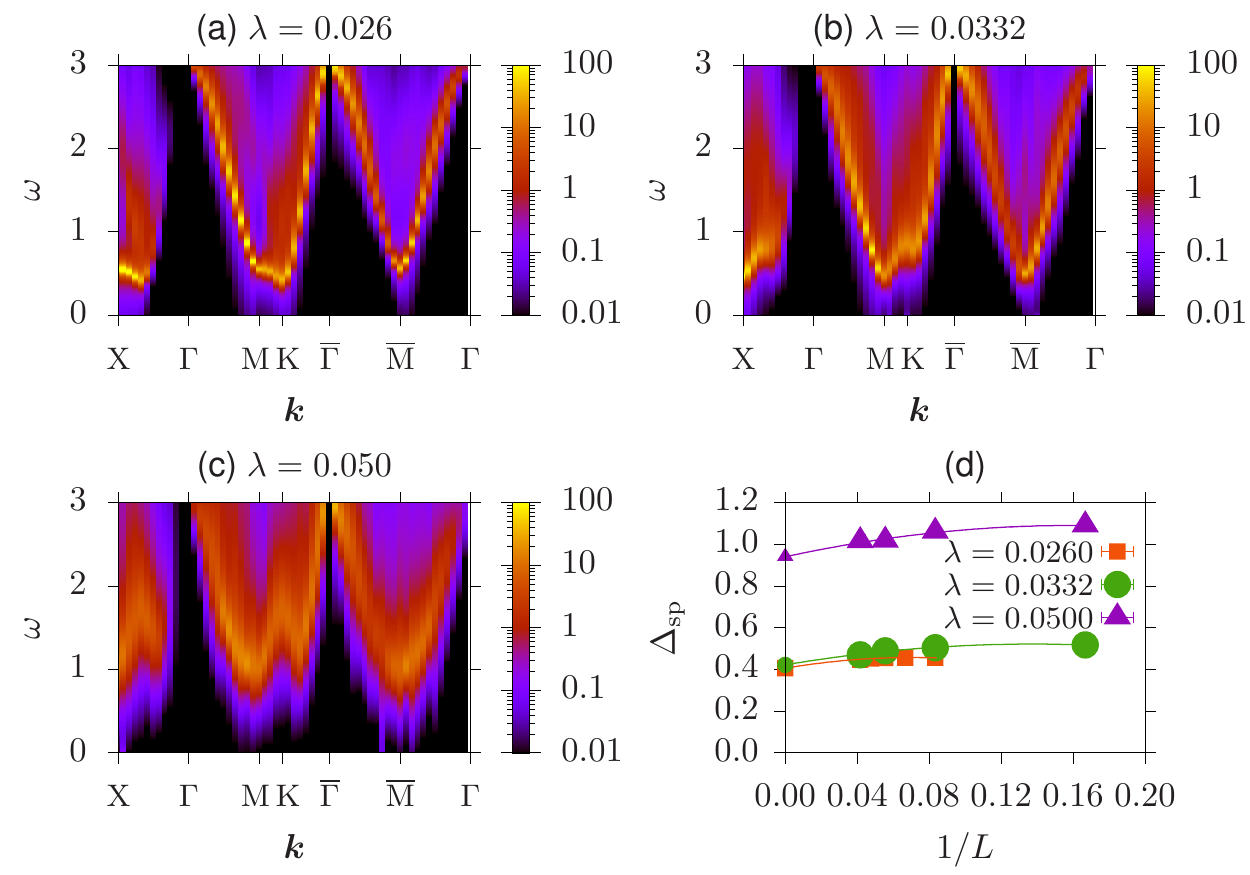}
  \caption{\label{fig:sp-green} (a)--(c) $T=0$ single-particle spectral
    function for $L=24$. (d) Finite-size scaling of the single-particle
    gap. The path in the Brillouin zone is shown in Fig.~\ref{fig:extended-zone}(c).}
\end{figure}

Figure~\ref{fig:sp-green} shows the single-particle spectral function at
$\lambda=0.026$, $\lambda=0.0332$, and $\lambda=0.05$.  As apparent, throughout
the considered parameter range, the single-particle gap remains finite.  
As discussed in Sec.~\ref{sec:model}, this leads to an emergent SO(3)$\times$U(1) symmetry, corresponding to rotations
in spin space and charge conservation, respectively.  Since the QSH order corresponds to a Dirac mass, the
single-particle gap in the vicinity of the DSM-QSH transition opens at the Dirac
point, K.  At $\lambda=0.026$, we see that the gap is indeed determined by the
Dirac point.  However, upon increasing $\lambda$, the position of the minimal
gap changes from the Dirac point K to the M point.

The $T=0$ single-particle gap $\Delta_\text{sp}$---defined as the
  minimal energy for single-particle excitations---can be extracted
by fitting the tail of the single-particle Green function at the corresponding
wave vector (K for $\lambda=0.026$, M for $\lambda=0.0332$ and $0.05$) to a
single exponential. The finite-size extrapolation shown in
Fig.~\ref{fig:sp-green}(d) reveals that the gap remains nonzero in the
thermodynamic limit.  These results are qualitatively
consistent with those obtained previously using a projective QMC method
\cite{liu2019superconductivity}, although in the former work $\Delta_\text{sp}$
was determined at K for all values of $\lambda$.

\subsubsection{QSH and SSC dynamical correlations}

\begin{figure}[t]
  \centering
  \includegraphics[width=0.48\textwidth]{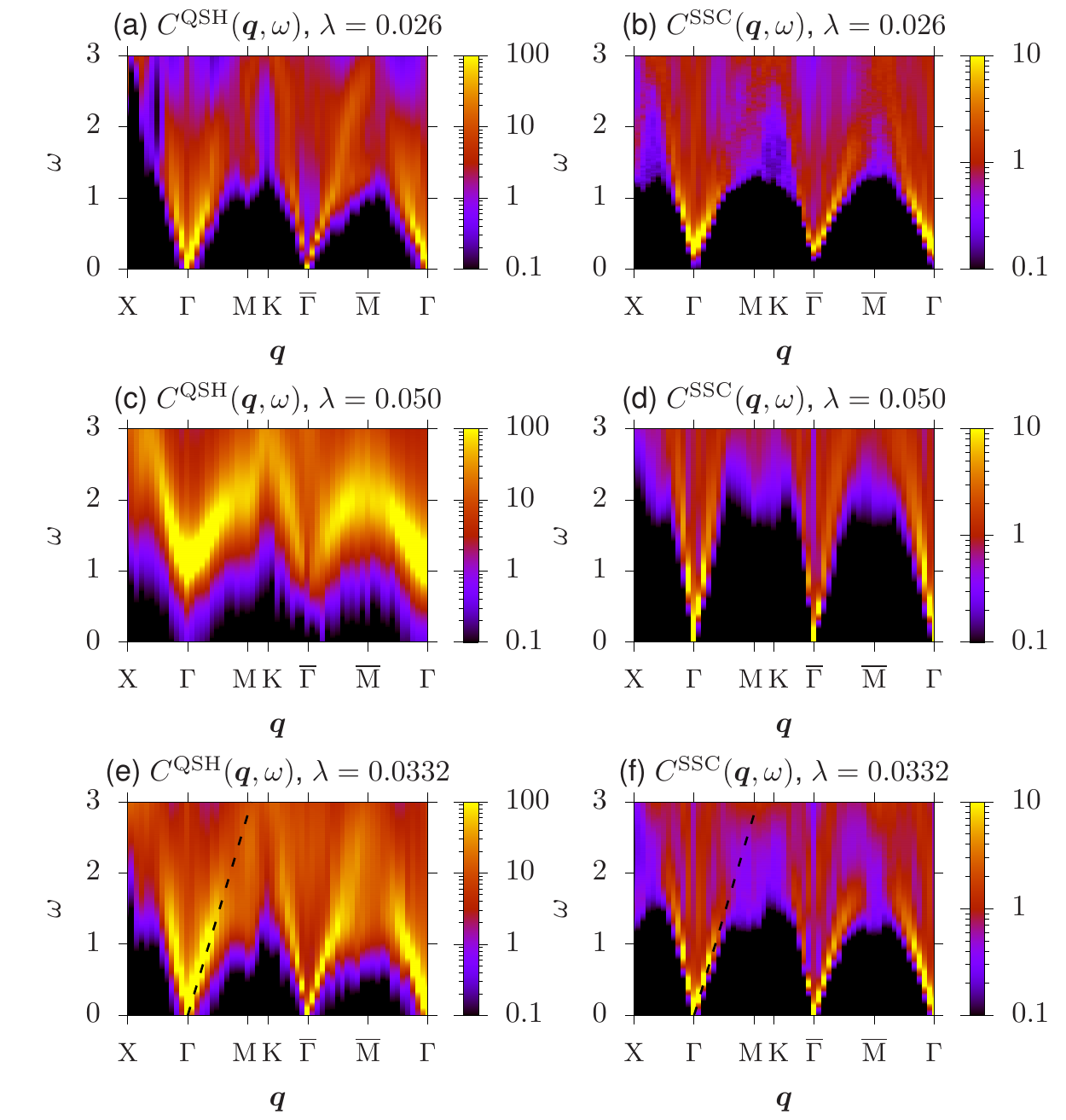}
  \caption{\label{fig:qsh-ssc-new}
    Zero-temperature dynamical QSH  [(a),(c),(e)] and s-wave pairing [(b),(d),(f)] structure factors
    at different values of $\lambda$ for $L=24$. The dashed lines in (e) and (f) correspond to $\omega= a
      (\ve{q}-\Gamma)$ with the same $a$. The path in the Brillouin zone is shown in Fig.~\ref{fig:extended-zone}(c).}
\end{figure}

In Fig.~\ref{fig:qsh-ssc-new}, we present the $T=0$ dynamic structure factors of the
QSHI and SSC order parameters  $(\alpha={\rm QSH, SSC})$,
\begin{eqnarray}
  \label{Cq}
C^{\alpha}(\boldsymbol{q}, \omega) = \pi
\sum_{n}  | \langle \Psi_n | \ve{\hat{O}}^{\alpha}_{\boldsymbol{q}}     | \Psi_0 \rangle |^2 \delta (E_n - E_0 - \omega)\,,
\end{eqnarray}
again calculated in the extended zone scheme.

The superconducting order parameter is a local quantity. Therefore, the
definition of $\hat{O}_{\ve{q}} $ in Eq.~(\ref{Eq:spectral}) is not ambiguous,
see also Fig.~\ref{fig:extended-zone}(a), and reads
\begin{equation}
  \hat{O}^\text{SSC}_{\ve{q}}   =   \frac{1}{\sqrt{V} }
  \sum_{\ve{r}} e^{i \ve{q} \cdot \ve{r}}  \left(   \text{Re}\left[
      \hat{a}^{\dagger}_{\ve{r},\uparrow}
      \hat{a}^{\dagger}_{\ve{r},\downarrow}  \right]
    +   \text{Re} \left[ \hat{b}^{\dagger}_{\ve{r},\uparrow}  \hat{b}^{\dagger}_{\ve{r},\downarrow}   \right]
    e^{i\ve{q} \cdot \ve{R}}   \right).
\end{equation}
In contrast, the QSH operator resides on links and we have used the mid-point of the
link to define the Fourier transform in the extended zone scheme:
\begin{eqnarray}
  \hat{\ve{O}}^{\rm{QSH}}_{\ve{q}}  = \frac{1}{\sqrt{V}} \sum_{\ve{r}}
  e^{i  \ve{q} \cdot \ve{r}} \sum_{\delta=1}^{6}\hat{O}_{\ve{r},\delta}^{\rm{QSH}}e^{i\ve{q} \cdot \ve{R}_{\delta}}.
\end{eqnarray}
Here, $\ve{r}$ corresponds to the center of the hexagon (there is a one-to-one
mapping between unit cell coordinates and the center of the hexagons). As shown
in Fig.~\ref{fig:extended-zone}(b),
$\ve{R}_{1}=( \boldsymbol{a}_2/2-\boldsymbol{a}_1)/3$,
$\ve{R}_{2}=( \boldsymbol{a}_2-\boldsymbol{a}_1/2)/3$,
$\ve{R}_{3}=( \boldsymbol{a}_2+\boldsymbol{a}_1)/6$,
$\ve{R}_{4}=-( \boldsymbol{a}_2/2-\boldsymbol{a}_1)/3$,
$\ve{R}_{5}=-( \boldsymbol{a}_2-\boldsymbol{a}_1/2)/3$,
$\ve{R}_{6}=-( \boldsymbol{a}_2+\boldsymbol{a}_1)/6$.

In the QSHI, the SO(3)  symmetry  is spontaneously broken, as apparent
from the low-lying Goldstone modes at the $\Gamma$ points in Fig.~\ref{fig:qsh-ssc-new}(a). Skyrmions of the SO(3)
order parameter carry charge 2e, $\rho(\ve{r})  =   \frac{2e}{4 \pi}
\ve{N}(\ve{r}) \cdot \partial_x   \ve{N}(\ve{r}) \times \partial_y
\ve{N}(\ve{r})$, where $\ve{N}(\ve{r})$  is the normalized QSH order parameter \cite{Grover08}.
Therefore, adding an s-wave  pair to the system  corresponds to a skyrmion
addition process or, equivalently, a monopole. Given their charge 2e
and the charge-conserving QSH order parameter, the skyrmions' dynamics is
revealed by $C^{\text{SSC}}(\ve{q},\omega)$ (Fig.~\ref{fig:qsh-ssc-new}(b))
and indicates that they are massive in the QSHI. The gap to the first
skyrmion excitation is smaller than twice the single-particle gap ($ 2
\Delta_\text{sp} \approx 0.8$), thereby  confirming that
they are the elementary charge excitations for $\lambda = 0.026$.

The SSC spontaneously breaks a U(1) symmetry and is hence also characterized
by Goldstone modes, as visible in Fig.~\ref{fig:qsh-ssc-new}(d) at the
$\Gamma$ points. Our understanding is that the SSC is the result of
skyrmion condensation.  A  distinct feature of this state is that the vortex
of the U(1) order parameter carries a spin-1/2 degree of freedom
\cite{Grover08}, a spinon. Pairs of spinons transform as an SO(3) vector and
can be observed in $C^{\text{QSH}}(\ve{q},\omega)$. In fact, even for $\lambda = 0.05$,
very far from the critical point,  $C^{\text{QSH}}(\ve{q},\omega)$
shows  states well below the  particle-hole continuum at $2
\Delta_\text{sp}\approx 1.9$ (Fig.~\ref{fig:qsh-ssc-new}(c)). We interpret this
as a signature of the gapped two-spinon
continuum of the SSC.

\subsubsection{Spin and charge dynamics}

For the spin dynamics, $C^S(\ve{q} , \omega)\equiv S(\ve{q},\omega)$, the operator $\hat{O}_{\ve{q}}$
in Eq.~(\ref{Eq:spectral}) corresponds to the vector
\begin{equation}
  \hat{\ve{S}}_{\ve{q}} = \frac{1}{\sqrt{V} } \sum_{\ve{r}} e^{i \ve{q} \cdot
    \ve{r}} \frac{1}{2}\left( \ve{a}^{\dagger}_{\ve{r}}  \ve{\sigma} \ve{a}^{}_{\ve{r}}   +
    \ve{b}^{\dagger}_{\ve{r}} \ve{\sigma} \ve{b}^{}_{\ve{r}}   e^{i\ve{q} \cdot \ve{R}}   \right)
\end{equation}
where we have adopted the same notation as in Eq.~(\ref{cq.eq}); $\ve{\sigma} $
is a vector of Pauli spin matrices and
$\ve{a}^{\dagger}_{\ve{r}} = \left( a^{\dagger}_{\ve{r},\uparrow},
  a^{\dagger}_{\ve{r},\downarrow} \right) $.  For the charge dynamics,
$C^N(\ve{q} , \omega)\equiv N(\ve{q} , \omega) $, $\hat{O}_{\ve{q}} $ in
Eq.~(\ref{Eq:spectral}) corresponds to
\begin{equation}
  {\hat{N}}_{\ve{q}}  =  \frac{1}{\sqrt{V} }   \sum_{\ve{r}} e^{i \ve{q} \cdot
    \ve{r}}
  \left( \ve{a}^{\dagger}_{\ve{r}} \ve{a}^{}_{\ve{r}}   +
  \ve{b}^{\dagger}_{\ve{r}}  \ve{b}^{}_{\ve{r}}   e^{i\ve{q} \cdot \ve{R}}   \right)\,.
\end{equation}

\begin{figure}[t]
  \centering
  \includegraphics[width=0.48\textwidth]{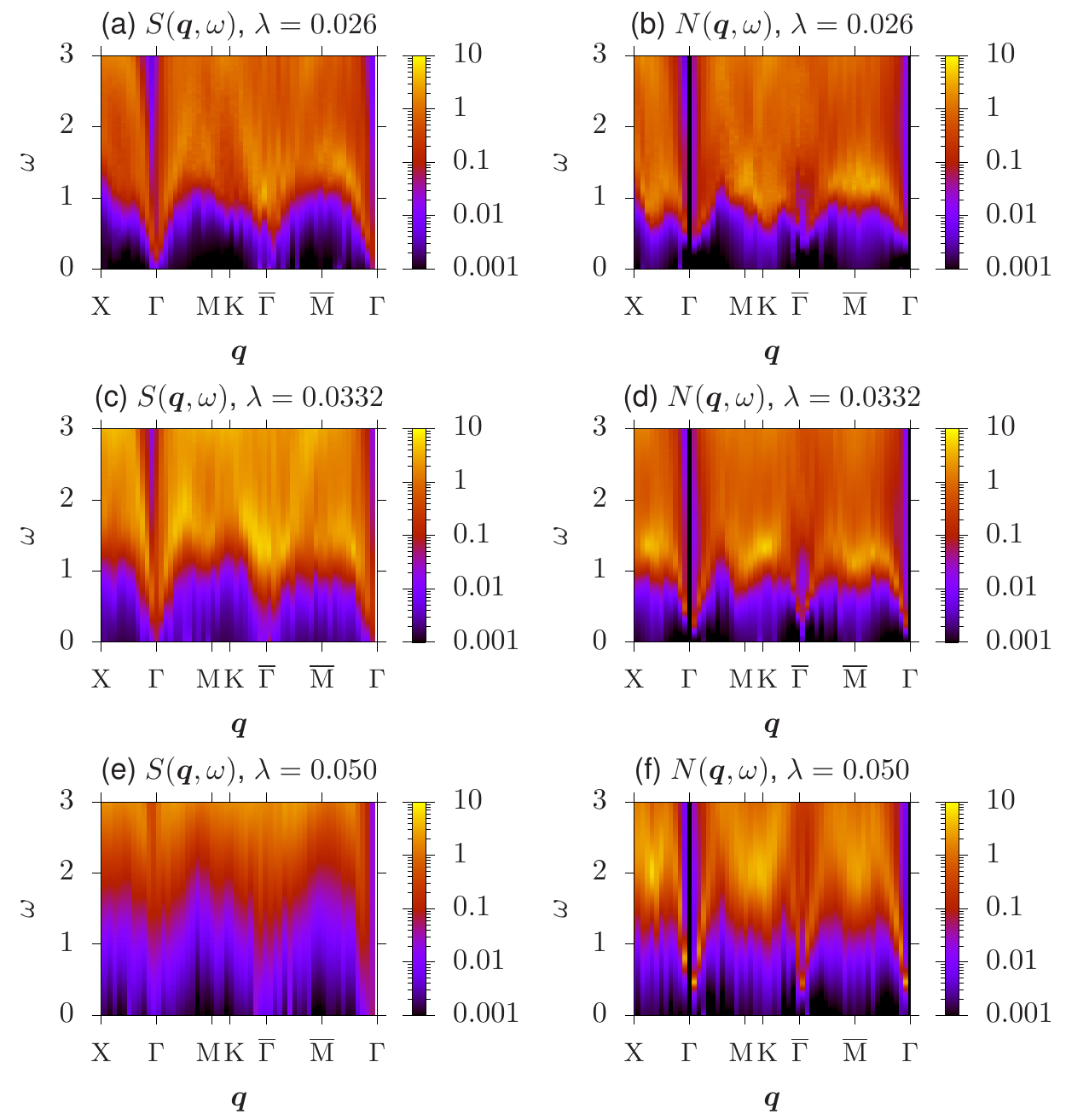}
  \caption{\label{fig:spin-charge} Zero-temperature spin [(a),(c),(e)]  and charge dynamic
    structure factors [(b),(d),(f)] for $L=24$. The path in the Brillouin zone is shown in Fig.~\ref{fig:extended-zone}(c)}
\end{figure}

The spin and charge dynamic structure factors at $\lambda=0.026$, $\lambda=0.0332$,
and $\lambda=0.05$ are shown in Fig.~\ref{fig:spin-charge}. The spin operator
transforms like the QSH order parameter under SU(2) rotations so that both
quantities probe spin excitations.  Similarly, both the superconducting order
parameter and the charge operator probe charge fluctuations.  We hence
expect to see similarities between the dynamical QSH (SSC) and dynamical spin
(charge) structure factors.  Aside from differences in spectral weight, and
accounting for uncertainties related to the analytical continuation,
this is indeed borne out by the results in Figs.~\ref{fig:qsh-ssc-new} and
\ref{fig:spin-charge}.  In particular, $S(\ve{q},\omega)$
($N(\ve{q},\omega)$) shows gapless (gapped) modes in the QSH phase,
Figs.~\ref{fig:spin-charge}(a),(b), whereas the behavior is reversed
in the SSC, see Figs.~\ref{fig:spin-charge}(e),(f).  At criticality,
Figs.~\ref{fig:spin-charge}(c),(d), both charge and spin dynamical responses
show very similar low-energy features. As pointed out in the main text, this
reflects an emergent SO(5) symmetry.

At $\ve{q}=\ve{0}$
(\ie, at $\Gamma$ and $\overline{\Gamma}$) the
spectral weight is expected to vanish because the structure factor probes the
total spin and total charge fluctuations.  These quantities correspond to the
generators of the spin and charge symmetries and vanish identically at zero
temperature. In our QMC simulations, the SU(2) spin symmetry is restored
only after the sampling (and, strictly speaking, in the limit $ \Delta_{\tau}
\to 0 $), so that the above statement is valid
only within error bars.  On the other hand, the charge is conserved for each
Hubbard-Stratonovich configuration.  This provides an explanation for why
$N(\ve{0},\omega) \equiv 0$ whereas $S(\ve{0},\omega)$ is small but finite.

Based on the above discussion, we expect the DQCP to be characterized by
gapless spin-1/2 spinons and gapless charge-2e skyrmions. This is confirmed
by Figs.~\ref{fig:qsh-ssc-new}(e) and (f). Note that at criticality, the
particle-hole continuum is located at $2 \Delta_\text{sp} \approx0.8$.
The gapless spin and charge excitations lead to the observed linear
temperature scaling of the charge and spin susceptibilities. The remarkable
similarity of the QSH and SSC dynamical structure factors below the
particle-hole continuum, most notably the identical velocities
of excitations indicated by the dashed lines, can be understood in 
terms of an emergent Lorentz  symmetry that implies the existence of a single velocity, the velocity  of light.

\section{Discussion}\label{sec:discussion}

Hamiltonian~(\ref{eq:hamiltonian1}) provides a realization of a DQCP between a QSHI
and an SSC.  The quantum critical region shown in Fig.~\ref{fig:phasediagram}(b)
has distinct thermodynamic and dynamical signatures: the single-particle gap is
finite but the spin and charge susceptibilities reveal gapless excitations in
the respective channels.  Gapless charge excitations capture skyrmions of the
SO(3) QSH order parameter with charge 2e.  Gapless spin excitations reflect the
vortices of the U(1) SSC order parameter with spin-1/2.  A natural upper bound
for the quantum critical region is the single-particle gap.  Below this
energy scale, the very symmetry of the quantum critical theory emerges. In fact,
the generation of quantum anomalies by gapping out fermions has been put forward
in Ref.~\cite{Thorngren20}, where it is conjectured that the state at
$\lambda_{c2}$ is an intrinsically gapless topological state.

The interpretation of the data proposed above naturally emerges when describing
the DQCP in terms of a five component non-linear sigma model with
SO(3)$\times$U(1) symmetry subject to a Wess-Zumino-Witten term
\cite{Tanaka05,Abanov00,Senthil06,Grover08}. This field theory is derived
perturbatively from Dirac fermions subject to a Yukawa coupling to the five
anticommuting QSH and SSC mass terms \cite{Ryu09} by integrating out the gapped
fermionic degrees of freedom.  There is evidence of an emergent
SO(5) symmetry at intermediate energy scales \cite{Nahum15_1,WangZ20}.  In
particular, the anomalous dimensions of the QSH and SSC order parameters are
equal within error bars \cite{liu2019superconductivity}.  It should be noted that, at
the energy scales set by our finite lattice sizes, we cannot distinguish between
weakly first-order transitions---characterized by the proximity to a critical
point with SO(5) symmetry \cite{WangC19,Nahum19}---and genuine second-order
phase transition. The proximity to an SO(5) critical point is
consistent with our spectra and the symmetry implies identical supports
for the  QSH and SSC dynamical structure factors at criticality.

The transition to the SSC at $T_c$ is particularly interesting close to the DQCP
at energy scales well below the single-particle gap.  At $T_c$, vortices of the
SSC order parameter unbind.  Since they carry a spin-1/2 degree of freedom, the
resulting state above $T_c$ should be understood as a liquid of spinons. Within
the theory of the DQCP \cite{Senthil04_1,Senthil04_2}, this gas of spinons is
described by a non-compact CP$^1$ field theory in its deconfined phase. The
spinons are captured by the spin susceptibility whereas the charge
susceptibility reveals the gauge-field flux corresponding to the charged
skyrmions. Remarkably, the nontrivial vortex core does not seem to
impact the temperature-driven transition to the normal phase. Instead, the
finite-size scaling analysis underlying the determination of $T_c$ yields a
clean crossing point when assuming the usual 2D XY exponents, see Fig.~\ref{fig:TcSSC}(a).

The QSHI-SSC DQCP considered differs in important aspects from AFM-VBS DQCPs in
JQ \cite{Sandvik07} or loop models \cite{Nahum15}. The JQ model has a second
length scale associated with the energy scale where the emergent U(1) symmetry
gives way to the C$_4$ point group symmetry of the
Hamiltonian.  This implies that the uniform spin susceptibility does
not scale linearly with $T$ at criticality, as confirmed numerically
\cite{Shao15}. 

Recently, doping of the QSH state at $\lambda = 0.026$ and $T=0$ was shown to
yield a continuous and direct transition to an SSC
\cite{wang2020dopinginduced}. The thermodynamic and dynamical properties of this
transition are of particular interest for future work.  The spectra in
Fig.~\ref{fig:qsh-ssc-new} reveal the existence of preformed pairs identified
here as skyrmions. The study of thermodynamic quantities at finite doping could
hence reveal pseudogap physics extensively discussed in connection with
high-temperature superconductivity.

\begin{acknowledgments}
  We thank Chong Wang for helpful discussions.  The authors gratefully
  acknowledge the Gauss Centre for Supercomputing e.V. (www.gauss-centre.eu) for
  funding this project by providing computing time on the GCS Supercomputer
  SUPERMUC-NG at the Leibniz Supercomputing Centre (www.lrz.de).  TS thanks the
  Deutsche Forschungsgemeinschaft (DFG) for funding via grant SA 3986/1-1.  FFA
  thanks the DFG for funding via grant AS~120/15-1 and the W\"urzburg-Dresden
  Cluster of Excellence on Complexity and Topology in Quantum Matter ct.qmat
  (EXC 2147, project-id 390858490). MH and ZW acknowledge support from the DFG
  via SFB 1170 ToCoTronics.  YL was supported by the China Postdoctoral Science
  Foundation under grants no.~2019M660432 and 2020T130046 as well as the
  National Natural Science Foundation of China under grants no.~11947232 and
  U1930402.  WG was supported by the National Natural Science Foundation of
  China under grants no.~12175015 and 11734002.
\end{acknowledgments}

\appendix*

\section{Scaling of conserved quantities}

We consider a Hamiltonian $ \hat{H} = \sum_{i} K_i \hat{O}_i $ that commutes
with the generator of a global symmetry, $\hat{G}$.  In our specific case,
$\hat{G}$ corresponds either to the total particle number or to a component of the
total spin. The partition function is block diagonal in $\hat{G}$,
\begin{eqnarray}
  Z\left( \ve{K}, L, \beta, \mu, l \right)
  &\equiv&
  \text{Tr}
  e^{- \beta \left(  \hat{H} -  \mu \hat{G} \right) }\\
  &=& \sum_{G} e^{\beta \mu G }
  Z_G\left( \ve{K}, L, \beta, l \right).
\end{eqnarray}
Here, $\mu$ corresponds to a Lagrange multiplier, $L$ is the linear length of
the system, $\beta$ the inverse temperature, $l$ an additional length scale and
$Z_G$ the partition function in the Hilbert space spanned by the vectors
$\hat{G} | \Psi \rangle = G | \Psi \rangle $.  We note that the additional
length scale also diverges at the critical point, albeit with a distinct
exponent.  The susceptibility reads
\begin{equation}
  \chi_G    =  - \left.\frac{ \partial^2 f}{ \partial \mu^2} \right|_{\mu = 0 }
  = \frac{\beta}{L^d} \left(  \langle  \hat{G}^2 \rangle  -
    \langle  \hat{G} \rangle^2 \right),
\end{equation}
where $f = -\frac{1}{\beta L^d} \log Z $ is the free-energy density and
$ \left< \hat{O} \right> = \frac{1}{Z} \text{Tr} \left[ e^{- \beta \hat{H} }
  \hat{O} \right] $. We carry out a renormalization group (RG) transformation in each $G$ sector:
\begin{equation}
  Z_G\left( \ve{K},   L, \beta,  l  \right)
  =  Z_G\left( \ve{K}^{(1)},   \frac{L}{b}, \frac{\beta}{b^z},  \frac{l}{b^{z_l}}  \right).
\end{equation}
Here, $z$ corresponds to the dynamical critical exponent and $z_l$ encodes the
presence of an additional length scale.  The RG transformations of
the couplings $\ve{K}$ and the exponents do not depend on $G$. This can be justified
as follows.  Consider a situation where there is a gap between the $G$ sector
corresponding to $\mu=0$ and all other sectors.  Then, fluctuations of $G$ are
exponentially suppressed as a function of temperature and it is meaningful to
consider a single $G$ sector.  In this case, $\chi_G $ vanishes exponentially
and hence exhibits no scaling behavior.  Let us now consider the more
interesting case where there is no gap in the $G$ spectrum.  In this case, we
expect that
$ { \sqrt{\langle\hat{G}^2\rangle - \langle \hat{G} \rangle^2 } }/ { \langle
  \hat{G} \rangle } $ vanishes in the thermodynamic limit.  We note that for,
say, $\hat{G}$ corresponding to the total particle number, the vanishing of the
latter relation implies that the particle number is sharp in the grand canonical
ensemble so that the canonical and grand canonical ensembles are equivalent in
the thermodynamic limit.  The same argument holds for the equivalence of the
canonical and micro-canonical ensembles.  Hence in the absence of a gap, $G$ is
still \textit{sharp}, in the above sense, so that an RG flow independent of $G$
is validated.

With this in mind, we can apply the RG transformation to the susceptibility and
obtain:
\begin{equation}
  \chi_G(\ve{K},L, \beta,  \mu, l  )
  =  \frac{b^z}{b^d}
  \chi_G\left(\ve{K}^{(1)} ,\frac{L}{b}, \frac{\beta}{b^z},  \frac{l}{b^{z_l}} \right).
\end{equation}
Let us now iterate the RG transformation $n$ times, place ourselves at the
critical point, $ \ve{K}^{(1)} = \ve{K} = \ve{K}^*$, set $b^n = L$,
$ \beta = L^z$ to get
\begin{equation}
  \chi_G(\ve{K}^*,L, \beta, l  )
  =  \beta^{1 - \frac{d}{z}} \chi_G\left(\ve{K}^* , \frac{l}{L^{z_l}}  \right)
  =    \beta^{1 - \frac{d}{z}}  f\left(  \frac{l^{1/z_l}}{L}  \right).
\end{equation}
Hence, in the absence of a second length scale, $z_l = 1 $, we conclude that
$ \chi_G(\ve{K}^*,L, \beta, l ) \propto \beta^{1 - {d}/{z}} $.  Given a second
length scale, corrections to this scaling law are expected.

%\bibliography{../fassaad}


\end{document}